\documentstyle[amssymb,prb,preprint,aps]{revtex}
\begin{document}
\draft
\title{Universal T$_{c}$ depression by irradiation defects in underdoped and
overdoped cuprates }
\author{F.\ Rullier-Albenque$^{1}$, P.A.\ Vieillefond$^{1}$, H.\ Alloul$^{1*}$, A.\
W.\ Tyler$^{2\times }$, P.\ Lejay$^{3}$ and J.F.$\;$Marucco$^{4}$}
\address{$^{1}$ Laboratoire des Solides Irradi\'{e}s, CEA, CNRS UMR 7642, Ecole\\
Polytechnique 91128\\
Palaiseau cedex, France, }
\address{$^{2}$ IRC in Superconductivity, University of Cambridge, Madingley Road,\\
Cambridge CB3 0HE, UK, }
\address{$^{3}$ CRTBT, CNRS, BP 166X, 38042 Grenoble,cedex,}
\address{$^{4}$ LEMHE,Universit\'{e} Paris-Sud, 91405\ Orsay cedex, France}
\date{\today}
\maketitle

\begin{abstract}
We report on a study of the influence of defects introduced in the CuO$_{2}$
planes of cuprates in a wide range of hole dopings $n$.\ T$_{c}$ and
electrical resistivity $\rho (T)$ measurements have been performed on
electron irradiated YBa$_{2}$Cu$_{3}$O$_{7-\delta }$ and Tl$_{2}$Ba$_{2}$CuO$%
_{6+x}$ single crystals. A universal scaling between the decrease in T$_{c}$
and $\Delta \rho _{2D}\times n$, where $\Delta \rho _{2D}$ is the increase
of the 2D-resistance induced by the defects, is found for all the samples
investigated here. This demonstrates that $n$ is the relevant parameter to
describe the transport properties all over the phase diagram, in
contradiction with a recent suggestion of a change in the number of carriers
from $n$ to $1-n$ at the optimal doping. Moreover, the analysis of our data
suggests that strong scattering persists on the overdoped side.

PACS. 74.62.Dh : Effects of crystal defects, doping and substitution

PACS. 74.25.Fy : Transport properties

PACS 74.72.-h : High T$_{c}$ compounds
\end{abstract}

%\tighten

It is well established that defects introduced in the CuO$_{2}$ planes of
cuprates alter very effectively the properties of these materials in both
their normal and superconducting states. In particular, the large T$_{c}$
depression induced by Zn, a ''non magnetic'' scatterer, has been extensively
studied \cite{Chien,Alloul,Maha,Mizu,Kluge,Fuku,Tallon1,Mendels}. NMR and
susceptibility measurements \cite{Alloul,Maha,Mendels} in Zn substituted
YBCO samples have shown that Zn induces localized moments on neighbouring Cu
sites. The size of these moments is tightly related to the hole doping,
decreasing from S$\sim $1/2 in underdoped samples to a very small value in
optimally doped ones. Concerning transport properties, Fukuzumi et al \cite
{Fuku} have used Zn substitution in single crystals of YBCO and LaSrCuO to
probe the evolution of the electronic state of high T$_{c}$ cuprates with
doping. Their results show a critical change in the normal state transport
properties from the underdoped to optimal or overdoped regime, with the
carrier density being identified with the density of doped holes in the
former regime and with the density of electrons in the latter one. Nagaosa
and Lee \cite{Naga} were able to account for these resistivity data in the
framework of spin-charge separation theory. They proposed that the crossover
in Zn induced residual resistivity increases should be due to a Kondo
screening effect - screening of localized spins by conduction electron spins
- which increases as the hole concentration increases and the local moment
disappears.

Electron irradiation is another practical way to introduce in-plane defects
in a controllable manner in high-T$_{c}$ cuprates \cite
{legris,Giapin,Tolpygo}. More precisely, it has been previously shown in YBCO%
$_{\symbol{126}7}$ that the irradiation defects relevant to explain the T$%
_{c}$ decrease are related to oxygen and copper displacements in the CuO$%
_{2} $ planes.\ They induce a T$_{c}$ depression quantitatively similar to
Zn substitution \cite{legris}. The great advantage of electron irradiation
is to allow transport measurements on the same sample with an increasing
concentration of in-plane defects ranging from 0 to several \%$\ $ without
changing the hole doping.

In this paper, we present results of resistivity measurements on single
crystals of optimally and under-doped YBCO$_{7-\delta }$ and optimally and
over-doped Tl$_{2}$Ba$_{2}$ CuO$_{6+\delta }$ (Tl-2201) irradiated at low
temperatures by 2.5 MeV electrons. This allows us to study the interplay
between in-plane defects and hole doping for a wide range of hole dopings.\
The remarkable result of this paper is to show that the relevant parameter
to describe the changes in T$_{c}$ and resistivity induced by in-plane
defects is the concentration of doped holes, in contradiction with the
analysis proposed in ref.6 and 9.

Single crystals of YBCO$_{7-\delta }$ were grown using the standard flux
method as described elsewhere \cite{lejay}. Some optimally doped crystals
were subsequently annealed in air at different temperatures and quenched at
room temperature to adjust their oxygen content. The evaluation of oxygen
content in our crystals was made by comparing T$_{c}$ values - taken as the
middle of the superconducting transition- with published data \cite{Ito}.
Three different crystals were studied with $\delta $ values of 0.07, 0.2 and
0.4.

Tetragonal single crystals of Tl$_{2}$Ba$_{2}$CuO$_{6+\delta }$ were grown
using a Cu-rich flux. Good electrical contacts are made by evaporating gold
pads onto the as-grown samples. Furthermore, the crystals are annealed using
various combinations of annealing temperature, atmosphere and time in order
to obtain T$_{c}$ values ranging from $\symbol{126}$2K to 83K \cite{Andre}.
Four different crystals with T$_{c}$ values of 80K, 54K, 33K and 31K have
been used.

The electron irradiations were performed in the low temperature facility of
the Van de Graaff electron accelerator at the LSI. During irradiation, the
samples are immersed in liquid H$_{2}$ (21K) and the irradiation flux is
limited to 2x10$^{14}$ e$^{-}$/cm$^{2}$ s in order to avoid heating of the
samples. In-situ resistivity measurements have been performed between 20K
and 300K.

Figures 1-a and 1-b show the temperature dependence of the in-plane
resistivity $\rho _{ab}$ for two single crystals of YBCO$_{7-\delta }$ with $%
\delta \sim $ 0.07 and $\sim $ 0.4 irradiated at 20K in the same run and
annealed up to 300K for increasing electron fluences. One observes the same
tendencies as usually found when defects are introduced into CuO$_{2}$
planes : T$_{c}$ shifts downwards and the absolute value of the resistivity
increases. For both optimal and underdoped compounds, the $\rho $(T) curves
are perfectly parallel showing that the irradiation defects induce a
temperature-independent contribution to resistivity, i.e. Matthiessen's rule
is very well verified.\ This also indirectly confirms that the hole doping
is not modified under irradiation in agreement with Hall effect data on
electron irradiated YBCO$_{7}$ \cite{legris}.\ Therefore one can determine
the increase of residual resistivity $\Delta \rho $ due to in-plane defects
with high accuracy as the vertical shift of the $\rho $(T) curves. In
underdoped crystals we can note that defects do not affect the
characteristic temperature below which the pseudo gap starts to open as
already observed \cite{Alloul,Mizu,Tolpygo}. Moreover a metal-insulator
transition is visible once resistivity values are around 300$\mu \Omega $%
.cm. We have also observed this transition in optimally doped YBCO crystals
at about the same resistivity value but for much higher electron fluences 
\cite{Alex}.

Intermediate measurements have been carried out without warming the samples
above 100K.\ In such conditions we have previously shown the absence of
defect annealing \cite{legris}. Very good linear dependences of the residual
resistivity and the critical temperature versus electron fluences are found
(fig.2). Similarly to the case of Zn substitution the decrease rate of T$%
_{c} $ is faster on the underdoped side, increasing by a factor $\sim $2
from O$_{7}$ to O$_{6.6}$ as reported in the inset of fig. 2.

In the case of YBCO$_{7}$, the threshold energies necessary to displace
oxygen and copper atoms have been previously determined to be equal
respectively to 10 and 15 eV \cite{legris}. The number of defects created
per CuO$_{2}$ plane is $n_{d}^{i}$ = $\sigma _{d}^{i}\varphi t$ where $%
\sigma _{d}^{i}$ is the cross-section for displacing the atom i from its
site and $\varphi t$ is the electron fluence.\ Calculations of $\sigma
_{d}^{i}$ \cite{lesueur} lead to a total number of displaced O and Cu atoms
per CuO$_{2}$ plane $n_{d}$ $\simeq $ 3.7$\times $10$^{-3}$ for an electron
fluence of 10$^{19}$ e/cm$^{2}$. Even if this absolute value of $n_{d}$ is
not very accurate, one does not expect a variation with the oxygen content.
This allows the precise comparison between the three different crystals
studied here which is illustrated in figure 2. The rate of increase of the
2D-resistivity $\Delta \rho _{2D}$ increases from 0.6 to 1.9 k$\Omega $
/\%defects with decreasing oxygen content from O$_{7}$ to O$_{6.6}$. To a
first approximation we can consider that defects related to oxygen and
copper displacements behave as identical scatterers \cite{note 1}. Given the
analogy emphasized above between irradiation defects and Zn substitution, it
is tempting to compare our results for $\Delta \rho _{2D}$ to those found
for Zn substitution. In YBCO$_{6.6}$\cite{Fuku}, $\Delta \rho _{2D}$ has
also been determined from the parallel shift of the $\rho (T)$ curves.\ The
obtained values of $\Delta \rho _{2D}/n_{d}$ are similar for Zn and
irradiation defects.\ However, in YBCO$_{\symbol{126}7}$ raw data reported
in ref. 1 and 6 display an increase of the slope of the $\rho (T)$ curves
with Zn substitution in contrast with our results (see fig. 1-a). This could
indicate that the Zn impurities introduce a temperature-dependent
contribution to resistivity.\ In view of our experiment, this explanation
seems unlikely and we rather think that this apparent breakdown of
Matthiessen's rule signals a difference in hole doping between the pure and
Zn substituted single crystals used by Fukuzumi et al\cite{Fuku}.\ As a
consequence the analysis in ref. 6 in which $\Delta \rho _{2D}$ is
determined as the 0K intercept of the T-linear part of $\rho (T)$ seems to
us questionable.\ 

If we assume as is usually done that the contribution of the s-wave channel
dominates the scattering processes, the increase in $\rho _{2D}$ can be
written as

\begin{equation}
\Delta \rho _{2D}=\frac{4\hslash }{e^{2}}\frac{n_{d}}{n}\sin ^{2}\delta ,
\label{equation1}
\end{equation}
where $\delta $ is the scattering phase shift and $n$ the carrier
concentration. We have plotted in fig. 2 the values of $\Delta \rho _{2D}$
obtained in the unitarity limit ($\delta $ = $\pi /2$) assuming that the
carriers are the doped holes (with concentration $x$), that is $n=x$. We
find that the lower the hole doping, the closest the experimental data with
these unitarity limit estimates. Here we have taken $x$ to be equal to 0.16,
0.12 and 0.09 respectively as the oxygen content decreases, using the
relationship between T$_{c}$ /T$_{c0}$ and $n$ proposed by Tallon et al \cite
{Tallon}. These values are somewhat low compared to those usually assumed in
the literature. Nevertheless, even with a value as high as $x=$ $0.23$ taken
by Fukuzumi et al. \cite{Fuku}, our experimental result for YBCO$_{7}$
cannot be accounted for by using equation (\ref{equation1}) with $n=1-x$ as
proposed in ref.6 and 9 for Zn substituted crystals.\ One can see in fig.2
that our data lie well above the straight line obtained in this limit. As
discussed above, the different conclusion in the case of Zn mainly arises
from the difficulty encountered in ref.6 to determine the increase of
residual resistivity.

Contrary to the case of YBCO, the response of Tl-2201 crystals to low
temperature irradiation strongly depends on the doping level of the samples.
The optimally doped crystals behave similarly to YBCO$_{7}$ : defects
introduced at 20K are stable up to $\sim $130K and the resistivity above T$%
_{c}$ exhibits a linear T dependence obeying Matthiessen's rule. However in
overdoped crystals, irradiation at 20K induces a very fast decrease of T$%
_{c} $ which the major part is recovered after warming up the samples to
80K. Meanwhile, the irradiation induced increase of the normal state
resitivity exhibits no thermal recovery. We think that this might be
associated with a change of hole doping induced by oxygen displacements in
the charge reservoir planes which are reversible after annealing above 80K.
A detailed analysis of these effects will be reported elsewhere \cite{Tl a
venir}. Nevertheless after systematic annealing at 300K, one can see in
figure 1-c that Matthiessen's rule is also very well verified for the most
overdoped Tl-2201 crystal studied here. This is for us the indication that
the residual defects remaining in the samples after this thermal treatment
do not modify the hole doping. One can therefore conclude that the prime
cause of T$_{c}$ suppression after annealing at 300K is again due to the
presence of defects in the CuO$_{2}$ planes. As for YBCO, the increase of $%
\Delta \rho $ due to these defects can be determined very precisely by the
parallel shift of the $\rho (T)$ curves.

In this condition the actual value of $n_{d}$ in the CuO$_{2}$ planes cannot
be estimated reliably. However in order to compare the effect of irradiation
in Tl-2201 and YBCO we have plotted in figure 3 the decreases in T$_{c},$ $%
\Delta $T$_{c}$, as a function of $\Delta \rho _{2D}$ (a) and $\Delta \rho
_{2D}\times n$ (b) with $n=x$ for all the samples studied here . We have
used the same approach as described above to estimate the hole doping in
Tl-2201 \cite{Ober}, which leads to $x$ = 0.16, 0.22 and 0.25 for the
optimally doped samples and for the overdoped ones with initial T$_{c}$ of
54 and 31K respectively. Data obtained on the crystal with initial T$_{c}$
of 33K give similar results as the one with T$_{c}$=31K and has not been
reported in fig.3.\ The striking result displayed in fig.3b is that,
irrespective of the compound or of the hole doping, the data are located on
the same pair-breaking curve. It is worth mentioning that such a scaling is
also followed by single crystals of Bi-2212 \cite{Bi} and lightly underdoped
single crystals of Hg-1223\cite{Pekin} irradiated in the same conditions.\ 

This universal scaling strongly suggests that $\Delta $T$_{c}$ is due to
impurity scattering in a d-wave superconductor as first suggested by Radtke
et al \cite{Radtke}.\ It is then given by the standard Abrikosov Gork'ov
formula which for small $n_{d}$ is written as

\begin{equation}
\Delta T_{c}=-\frac{\pi }{4k_{B}}\Gamma _{n},  \label{equation2}
\end{equation}
where $\Gamma _{n}=$ $\frac{n_{d}}{\pi N(E_{F})}\sin ^{2}\delta \ $ is the
scattering rate in the normal state and $N(E_{F})$ the density of states at
the Fermi level. \cite{Borko,Fehren}.\ With a cylindrical Fermi surface,
equation (\ref{equation2}) transforms into : 
\begin{equation}
\Delta T_{c}=-\frac{\pi e^{2}\hbar }{8k_{B}m^{*}}\Delta \rho _{2D}\times n.
\label{equation3}
\end{equation}
As comparable values of the effective mass m$^{*}$ are found for YBCO and
Tl-2201\cite{note2}, our experimental results give strong support to eq.(\ref
{equation3}) and to a d-wave symmetry of the pairing state all over the
phase diagram. It is worth noting here that an$\ $analysis in terms of
anisotropic s-wave superconductors would lead to a similar expression as eq.(%
\ref{equation3}) with the degree of anisotropy $\chi $ (0$\leq \chi \leq 1)$
as a proportionality coefficient in eq.(\ref{equation3})\cite{Tolpygo,openov}%
. Such an analysis would be unlikely as this parameter $\chi $ can hardly be
the same for the two compounds studied here whatever their hole dopings.
Moreover the correlation displayed in fig.3 is for us the clear indication
that $x$ is the relevant parameter to describe transport properties in the
cuprates. We have also plotted in this figure the data of ref.6 on Zn
substituted YBCO and LaSrCuO single crystals.\ To take into account our
criticism of the determination of $\Delta \rho $ in the case of YBCO$_{7}$,
we have estimated the residual resistivity from the data at 150K. One can
see that the corresponding data for underdoped and optimally doped YBCO are
consistent with the pair-breaking curve for irradiation defects. Such a
result does not apply for the La$_{1-x}$Sr$_{x}$CuO$_{4}$ system.\ One
explanation can be found in the observation that the increase in resistivity
due to Zn substitution strongly depends on the initial state of the samples
as recently shown for La$_{0.75}$ Sr$_{0.15}$CuO$_{4}$ films \cite{Marta}.\
On the other hand, several experiments such as thermopower or NMR
measurements \cite{Ober,Bobroff} have shown that this system exhibits
singular properties which could indicate that it is not a generic compound
for the cuprates.

Finally let us address the question of the evolution of sin$^{2}\delta $
with doping.\ The scaling displayed in fig. 3b indicates that $\Delta $T$%
_{c} $ per in-plane defect is proportional to sin$^{2}\delta $. This is
quite compatible with our analysis of $\Delta \rho _{2D}$ shown in fig.2.\
Indeed the decrease of $\Delta $T$_{c}/n_{d}$ at increasing doping can be
correlated to a decrease of $\delta $ from the unitarity limit $\delta =\pi
/2$.\ On the overdoped side we have no direct estimate of sin$^{2}\delta $.
Studies of impurities substitution in different overdoped cuprates have
shown that $\Delta $T$_{c}/n_{d}$ decreases up to $n$ $\sim 0,2$ and then
remains constant \cite{Kluge2,Tallon1}.\ If such a behaviour applies for the
Tl-2201 system, the correlation reported in fig. 3 suggests that the phase
shift goes on decreasing in the overdoped region.\ Taking similar values of
sin$^{2}\delta $ for optimally doped YBCO and Tl-2201 leads to a crude
estimate sin$^{2}\delta \sim 0.4$ for $n$ $\sim 0.25\ $ which suggests that
strong resonant impurity scattering persists on the overdoped side. Although
the origin of this resonant scattering with realistic potential strength is
not clear at the present time, this indicates strong correlations in the
host system as proposed on the basis of a t-J model \cite{poilblanc}.\ 

In summary, the analysis of resistivity data in electron irradiated YBCO and
the universal scaling found between $\Delta $T$_{c}$ and the pair-breaking
parameter $\Gamma _{n}$ show unambiguously that the number of doped holes is
the relevant parameter to describe transport properties in cuprates from the
underdoped to the overdoped regime. Moreover, the large estimated scattering
phase shift suggests strong scattering, which could result from large enough
electronic correlations even in the overdoped regime.

$^{*}$also at Laboratoire de Physique des Solides, Universit\'{e} Paris XI,
91405 Orsay cedex, France

$^{\times }$Present adress : School of Physics and Astronomy, University of
Birmingham, Edgbaston, Birmingham B15 2TT, United Kingdom

\newpage

\begin{figure}[tbp]
\caption{ Temperature dependences of the in-plane resistivity in YBCO$_{7}$
(a), YBCO$_{6.6}$ (b) and overdoped Tl-2201 (c).\ The $\rho (T)$ curves have
been recorded after systematic annealing at 300K.\ The two YBCO crystals
have been irradiated in the same run with fluences of 0, 2.5, 3.8 and 6 $%
\times 10^{19}$ e/cm$^{2}$ respectively for curves 0 to 3.\ In the case of
overdoped Tl-2201, the electron fluences were 0, 0.8 and 3.4 $\times 10^{19}$
e/cm$^{2}$.\ In the case of irradiated Tl-2201, the appearance of an
anomalous $\rho (T)$ peak just above the transition probably results from an
non-uniform current density distribution in this specific crystal%
\protect\cite{Siebold}.}
\label{fig1}
\end{figure}

\begin{figure}[tbp]
\caption{Increase of the 2D resistance measured at 100K (without warming the
samples above 100K) for the different YBCO$_{7-\delta }$ single crystals as
a function of the electron fluence and the in-plane defect concentration $%
n_{d}$.\ The solid lines indicate the fit using eq.(\ref{equation1}) in the
unitarity limit with the indicated values for the carrier concentration $n$
.\ Inset : Corresponding decreases of the critical temperature versus $%
n_{d}. $}
\label{fig2}
\end{figure}

\begin{figure}[tbp]
\caption{Decrease in the critical temperature $\Delta T_{c}$ as a function
of : (1) the increase in 2D resistance $\Delta \rho _{2D}$ (measured at
100K) and (2) the quantity $\Delta \rho _{2D}\times n$ for the different
samples studied ($\circ $ YBCO$_{7},$ $\bullet $ YBCO$_{6.8},$ $\times $ YBCO%
$_{6.6,}$ $\square $ optimally doped Tl-2201, $\boxplus $ and $\blacksquare $
overdoped Tl-2201 with initial T$_{c}$ of 54 and 31K. The empty and closed
diamonds are the results for Zn-substituted YBCO$_{7-\delta }$ crystals with 
$\delta =$ 0.07 and 0.37 with resistivity data taken at 150K. The empty and
closed triangles are for Zn-substituted La$_{2-x}$Sr$_{x}$CuO$_{4}$ with x =
0.15 and 0.2 with resistivity data taken at 50K\protect\cite{Fuku}.}
\label{fig3}
\end{figure}


\begin{references}
\bibitem{note 1}  As far as YBCO$_{\symbol{126}7}$ is concerned, we have
been able to determine the contribution to $\Delta \rho $ of the oxygen and
copper defects.\ These two values are not far from each other, which allows,
in a first approximation, to consider them as a ``mean defect''.
\end{references}
\end{document}